\documentclass[12pt]{article}

\usepackage{amsmath}
\usepackage{epsfig}

\makeatletter
\renewcommand\section{\@startsection {section}{1}{\z@}%
                                   {-3.5ex \@plus -1ex \@minus -.2ex}%
                                   {2.3ex \@plus.2ex}%
                                   {\normalfont\large\bfseries}}
\renewcommand\subsection{\@startsection{subsection}{2}{\z@}%
                                     {-3.25ex\@plus -1ex \@minus -.2ex}%
                                     {1.5ex \@plus .2ex}%
                                     {\normalfont\normalsize\bfseries}}
\renewcommand\subsubsection{\@startsection{subsubsection}{3}{\z@}%
                                     {-3.25ex\@plus -1ex \@minus -.2ex}%
                                     {1.5ex \@plus .2ex}%
                                     {\normalfont\normalsize\bfseries}}
\renewcommand\paragraph{\@startsection{paragraph}{4}{\z@}%
                                    {3.25ex \@plus1ex \@minus.2ex}%
                                    {-1em}%
                                    {\normalfont\normalsize\bfseries}}
\makeatother

\newcommand\as{\alpha_s}

\begin{document}

\begin{center}
 \LARGE \sf{Geometric Scaling in a Symmetric Saturation Model}%
{\renewcommand{\thefootnote}{\fnsymbol{footnote}}\footnote{
This research was partially supported by the EU Framework TMR
 programme, contract FMRX-CT98-0194.}}
\end{center}
\vspace*{0.1cm}
\begin{center}
{\sc  St\'ephane Munier}
\end{center}
\vspace*{0.1cm}
\begin{center}
{\it
Universit\`a di Firenze\\
Dipartimento di Fisica,
via G. Sansone 1\\ 
50009 Sesto Fiorentino,
Italy}\\
{\tt munier@fi.infn.it}
\end{center}
\vspace*{0.4cm}

\begin{abstract}
We illustrate 
geometric scaling
for the photon-proton cross section with a very
simple saturation model. 
We describe
the proton structure function
$F_2$ at small-$x$ in a wide kinematical range with an 
elementary functional form and a
small number of free parameters.
We speculate that the symmetry between low and high $Q^2$ 
recently discovered in the data
could be related to a well-known
symmetry of the two-gluon-exchange dipole-dipole
cross section.
\end{abstract}

\section{Introduction}

Golec-Biernat and W\"usthoff (GBW) have shown \cite{Golec-Biernat:1998js}
that a saturation model is able to describe the HERA data for the
proton structure function $F_2(x,Q^2)$
at small values of Bjorken variable $x$. 
The agreement of their model with
the data is quite good for all available values of $Q^2$, including
the lowest ones. 

The GBW model is based on a dipole picture
of photon-proton interaction \cite{Nikolaev:1991ja}.
Its main ingredient is that at high energy, a particular frame
can be chosen in which the interaction factorizes in two processes
well separated in time. First the photon fluctuates in a $q\bar q$ pair
of given transverse size $r_\gamma$, which then scatters
off the proton. The factorized dipole-proton cross section 
$\sigma^{dp}$ contains
the QCD evolution. It only depends on the size of the dipole
$r_\gamma$ and on the Bjorken variable $x$.
The following parametrization
was chosen by GBW:
\begin{equation}
\sigma^{dp}(x,r_\gamma)
=\sigma_0\left(1\!-\!\exp(-r_\gamma^2Q_s^2(x)/4)\right)\ ,
\label{eq:gbwmod}
\end{equation}
where it was assumed that the transverse momentum scale $Q_s$ had
a power-like dependence on $x$:
\begin{equation}
Q_s^2(x)=1\ \mbox{GeV}^2\cdot\left(\frac{x}{x_0}
\right)^{-\lambda}\ .
\label{eq:qs}
\end{equation}
This choice was motivated by the following considerations.
When $rQ_s/2\!\ll\! 1$, the
model reduces to colour transparency. When one approaches
the region $rQ_s/2\!\sim\! 1$, the exponential in Eq.~(\ref{eq:gbwmod})
takes care of resumming
many gluon exchanges, in a Glauber-inspired way.
Intuitively, this is what happens when the proton
starts to look dark.
In the region $r Q_s/2\!\ge\! 1$, such a microscopic interpretation
is no more valid but the exponential
has the virtue to force the cross section 
to tend to a constant at large $r Q_s$
(which also means at small values of $x$ because of the $x$-dependence
of $Q_s$),
thus respecting the unitarity constraint through a strict compliance
with the Froissart bound.
These formulae
involve 3 free parameters only, $\sigma_0$, $x_0$ and
$\lambda$. 
Once those are determined by a fit to the $F_2$ data,
one can take advantage of the universality of
the dipole cross section to
extend the model to more exclusive processes. Diffractive
structure functions \cite{Golec-Biernat:1999qd}
are predicted in a quite satisfactory way.
Elastic electro- and photoproduction of vector mesons
is also accounted for very successfully \cite{Caldwell:2001ky}.
Some attempts were recently made to predict the 
$\gamma^*-\gamma^*$ cross
section \cite{Timneanu:2001bk}.

However, several points remain unsatisfactory in this model,
both on the theoretical and on the phenomenological side.
First, the form of the dipole cross section is 
quite {\it ad hoc}.
Then, we note that for the description
of small-$Q^2$ data, an arbitrary
quark mass of order $140\ \mbox{MeV}$ has
to be chosen.
On the phenomenological side, the original saturation model fails
at describing the $\log 1/x$-slope of $F_2$ at large $Q^2$.
A revisited version has just been released \cite{Bartels:2002cj}, which
takes care of some
collinear resummations. The new model is much more successful,
but it seems to incorporate less saturation effects
since the saturation scale $Q_s$ is shifted towards a lower value
\cite{thorne}.
This might indicate that either saturation effects 
are there but can be
mimicked by a DGLAP evolution, or DGLAP effects 
are still predominant. In the latter case 
the success of the original dipole model
could be attributed to collinear resummations
effectively
taken into account in the dipole-proton cross section,
though this is not at all apparent.

Our goal in this paper is to go back to the main known features
of saturation and to incorporate them in a simple parametrization
for the structure function. 
Our main result is that an exponentiated elementary
dipole-dipole cross section (see
Eqs.~(\ref{eq:finalehighq2},\ref{eq:finalelowq2})
and Figs.~2,3,4)
fits remarkably well the HERA data.
We give some motivation and interpretation
for this parametrization in 
Sec.~\ref{sec:form} and we present the
comparison to the data in Sec.~\ref{sec:data}.

\section{Formulation}
\label{sec:form}

Following GBW,
we choose the so-called dipole-frame, in which
all the QCD evolution is incorporated in the proton.
The photon has just time to fluctuate in a $q\bar q$ pair
which subsequently interacts with a highly evolved
proton.
We describe qualitatively the energy evolution
of the proton in subsection~(\ref{subsec:sat}) and 
the scattering in Secs.~(\ref{subsec:s1}) 
and~(\ref{subsec:s2}). The model is summarized
on Fig.1.

\subsection{The saturated proton}

\label{subsec:sat}

When the center-of-mass energy increases (or equivalently
when the Bjorken variable $x$ decreases at fixed $Q^2$),
more and more quantum fluctuations are revealed in the proton.
Indeed, $x$ coincides with the fraction of longitudinal momenta 
(with respect to the proton momentum), 
of the probed partons.
The lifetime of a partonic fluctuation
is proportional to this quantity.
Thus going to smaller $x$ means probing shorter 
time intervals, and thus becoming sensitive to
more fluctuations.

The rise of the parton densities with $1/x$ is predicted 
in QCD
for a heavy onium \cite{Mueller:1994rr}: 
the density $\tilde n(x,k)$ of gluons of longitudinal
momentum fraction $x$ and transverse momentum $k$ 
is found to obey a linear integro-differential
equation, the BFKL equation \cite{Balitsky:1978ic,Kuraev:1977fs}.
Its solution shows that the gluon density increases
like a power of $x$:
$\tilde n(x,k)\sim x^{-\lambda}$.
Appart from a diffusion term which would appear in a more
refined treatment, this rise is independent of $k$.
This model is not correct for very small $x$
because the gluons interact 
with each other
and thus each energy level
can only accomodate a finite number of them:
the density of partons
of given transverse momentum  should saturate at some point.
This feature has been taken into account by a modification
of the BFKL equation.
This modification consists in supplementing it with a nonlinear
term:
\begin{equation}
\frac{\partial\tilde n(x,k)}{\partial\log(1/x)}
=\frac{\as N_c}{\pi}K\cdot\tilde n(x,k)-
\frac{\as N_c}{\pi} {\tilde n}^2(x,k)\ ,
\label{eq:sat}
\end{equation}
where $K$ is the linear BFKL kernel.
This is the GLR equation \cite{Gribov:1983tu}.
Several groups have derived saturation equations within QCD
(for a review, see Ref.~\cite{Iancu:2002xk, Mueller:2001fv}).
Among them, Kovchegov was able to write a simple equation
\cite{Kovchegov:1999yj} which reduces to
Eq.~(\ref{eq:sat}) when the partons are probed
by a dipole of small size.
All available approaches
to saturation predict that the transverse momenta of
the partons are on average shifted to a scale $Q_s$
called the saturation scale such that
\begin{equation}
Q_s^2(x)=\Lambda^2 e^{\lambda\log(1/x)}\ .
\label{eq:qsat}
\end{equation}
This was confirmed by a
numerical study presented in Ref.\cite{Golec-Biernat:2001if}.
Corrections to formula~(\ref{eq:qsat}) have been computed recently,
see Ref.\cite{Mueller:2002zm}.
The curve $Q^2\!=\!Q_s^2(x)$ in the $(x,Q^2)$ plane is
called the critical line.
The saturation scale
$Q_s(x)$ is believed to reach 1 GeV around $x\sim 10^{-4}$:
this is what comes out of the GBW phenomenological approach
but a direct experimental evidence was also provided 
in Ref.\cite{Munier:2001nr}.

It is convenient to pair the gluons in colour dipoles
of given size $r$
(which is possible in the large $N_c$ limit \cite{Mueller:1994rr}).
The dipoles are then characterized by a density $n(x,r)$.
$1/Q_s(x)$ is their average size. It is also
the mean distance between the center of neighbouring dipoles.
We shall assume that the evolved proton
is a collection of independent dipoles at the
time of the interaction whose sizes are distributed
around $1/Q_s(x)$.
We will approximate the distribution $n(x,r)$ by a delta function
of the mean value of the dipole size $\langle r\rangle\!=\!1/Q_s$.
Its complete evaluation is presently 
out of reach.
Our confidence in this approximation 
relies on the observation made in Ref.\cite{Stasto:2000er}
where it was shown that the data for 
$\sigma^{\gamma^*p}$ for any $x\!\leq\!10^{-2}$ and
$Q^2$ is function of the combined variable $Q/Q_s(x)$ only, to a
very good accuracy.

To fix the normalization of $n$, we assume that 
the total area of these
dipoles over the area of the proton is a constant ratio,
of order 1.
This is a kind of boundary condition
on the critial line, in the spirit
of Refs.\cite{Iancu:2002tr} and~\cite{Kwiecinski:2002ep}
and yields the constraint
\begin{equation}
\int\frac{d^2r}{r^2}\,n(x,r)=\frac{Q_s^2(x)}{\Lambda^2}\ .
\label{eq:n}
\end{equation}
This assumption is supported {\em locally} 
by the saturation equations of the type~(\ref{eq:sat}).
However, it also means implicitly
that throughout the evolution,
the area of the proton remains of order $1/\Lambda^2$.
This can only result from non-perturbative confinement effects.
It was indeed shown \cite{Kovner:2001bh} that 
the Kovchegov equation, i.e. local saturation 
alone would lead to violations of
unitarity due to the fast expansion of the
nucleon radius.
Note that the same hypothesis 
of a proton of fixed transverse size
is contained in GBW model.

The dipole distribution in the proton 
eventually reads
\begin{equation}
n(x,r)=\frac{Q_s^2(x)}{\Lambda^2}\cdot
2\pi r\cdot\delta(r\!-\!1/Q_s(x))\ .
\label{eq:approxprot}
\end{equation}

\subsection{The scattering}

\label{subsec:s1}

In our model where the proton is represented by a density $n$
of dipoles,
the cross section for
dipole-proton scattering reads
\begin{equation}
\sigma^{\gamma^*p}(x,Q^2)=\int {d^2r_\gamma}
\int_0^1 dz|\psi_Q(r_\gamma,z)|^2
\int \frac{d^2r}{r^2}\,
n(x,r)\cdot\sigma^{dd}(r_\gamma,r)\ ,
\label{eq:section}
\end{equation}
where $\psi_Q$ is the photon wave function on
a $q\bar q$ dipole state.

$\sigma^{dd}(r_\gamma,r)$ is the dipole-dipole cross section
computable in perturbative QCD. We only have to consider its
lowest order expression, since all the QCD evolution is put
in the dipole density $n$: there is no more room for radiative
corrections in $\sigma^{dd}$. It reads~\cite{Mueller:1994jq}
\begin{equation}
\sigma^{dd}(r_\gamma,r)=2\pi\as^2 r_<^2\left(1+\log\frac{r_>}{r_<}\right)\ ,
\label{eq:elem}
\end{equation} 
where $r_<=\min({r_\gamma,r})$ and $r_>=\max({r_\gamma,r})$.
The different factors that appear in this expression
have a straightforward interpretation.
This cross-section is the imaginary part of the dipole-dipole elastic
amplitude, which requires the exchange of at least two gluons
paired in a colour singlet.
These gluons must be able to resolve the smallest dipole, hence
their transverse momentum is larger than $1/r_<$.
This is colour transparency. It accounts for the factor
$r_<^2$.
The correcting logarithm, which becomes sizeable when the dipoles have
very different sizes,
comes from the monopolar Coulomb field,
which results in 
a logarithm in 2 dimensions. Indeed, when $r_>\!\gg\!r_<$, 
the smallest dipole sees
the constituents of the largest one as
two well-separated colour charges.

In order to have an equal treatment for the photon and the proton
and to get the simplest model, we treat the photon as a dipole
of size $1/Q$ instead of taking account of the complete 
probability distribution 
$\Phi(r_\gamma)\!=\!\int dz\, r_\gamma^2|\psi_Q(r_\gamma,z)|^2$.
This picture is relevant for a longitudinal photon
whose wave function is sharply peaked (around $\log r_\gamma\sim
\log(1/Q_s(x))$).
It is less relevant for transverse photons since their wave function
develops a plateau at large $Q^2$ 
\cite{Nikolaev:1994bg}: this is due to the well-known
fact that large size dipole configurations (aligned jet
configurations)
are always present in the transversely polarized photon.
We shall however ignore this difficulty. Indeed we believe it is not
essential to our discussion since anyhow we donot pretend to
be able to describe the very large $Q^2$ region.
In this spirit, we
approximate the distribution $\Phi$ by a Dirac distribution:
\begin{equation}
\Phi(r_\gamma)=
N\cdot 2\pi r_\gamma\cdot\delta(r_\gamma\!-\!1/Q)\ .
\label{eq:approxphot}
\end{equation}
$N$ is a normalization factor
that gives the rate of such a fluctuation
$\gamma^*\rightarrow q\bar q$.

We take $\as$ fixed and at the scale 1~GeV.
The approximation made here is that we stick to a 
kinematical region close
to $Q_s$ where $|\log(Q/Q_s)|\ll\log(Q_s/\Lambda)$.
Injecting Eqs.(\ref{eq:approxphot}), 
(\ref{eq:approxprot}) and (\ref{eq:elem}) into 
Eq.(\ref{eq:section}), we obtain:
\begin{align}
\label{eq:sansresompert}
\sigma^{\gamma^*p}(x,Q^2)&=N\frac{1}{\Lambda^2}
\frac{Q_s^2(x)}{Q^2}\cdot 2\pi\as^2
\left(1\!+\!\log\frac{Q}{Q_s(x)}\right)
\ \ \mbox{for}\ Q>Q_s(x)\\
\sigma^{\gamma^*p}(x,Q^2)&=N\frac{1}{\Lambda^2}
\cdot 2\pi\as^2
\left(1\!+\!\log\frac{Q_s(x)}{Q}\right)
\ \ \mbox{for}\ Q<Q_s(x)\ .
\label{eq:sansresom}
\end{align}
These formulae deserve an immediate interpretation.
When $Q>Q_s(x)$ (first formula), 
we are in the usual picture of DIS
where a photon of small size probes the parton (dipole)
content of the proton. The flux factor is $N$,
the target density is $Q_s^2(x)/\Lambda^2$, and the 
elementary cross section
is $2\pi\as^2(1\!+\!\log(Q/Q_s(x))/Q^2$.
This is the hard Pomeron regime: the cross section
grows like the exponential of the rapidity, due to
the multiplication of dipoles in the proton.

When $Q<Q_s(x)$ instead, the dipoles of the proton are smaller than
the dipole of the photon. 
The use of a perturbative elementary dipole-dipole cross section
is justified if $x$ is very small so that the scale of the
coupling constant $Q_s(x)$ is large enough. How large
has always been subject to debate. We will assume that
$Q_s\sim 1\ \mbox{GeV}$ is fine.
The proton looks like a collection of ``small'' size probes.
The picture is inverted with respect to the previous case:
$1/Q_s(x)$, which is both the mean dipole
size and their separation,
fixes the resolution.
The flux factor is $Q_s^2(x)/\Lambda^2$,
the target density is $N$ and the elementary dipole-dipole
cross section
is $2\pi\as^2(1\!+\!\log(Q_s(x)/Q))/Q_s^2(x)$.
The growth of the parton densities with the
energy is compensated by the photon-parton cross section,
which falls with the energy.
We are in the soft Pomeron regime, in which the
$\gamma^*-p$ cross section grows only like
a small power of the energy, $(W^2)^{0.08}$.
We note that in this region, the precise relationship between
the photon virtuality $Q^2$ and the dipole size is not very
relevant: corrections to it come as an additional 
energy-independent $\log$ term.

An important propery of this parametrization is first that it only
depends on $Q/Q_s(x)$ and second that the quantity
$Q/Q_s(x)\,\sigma^{\gamma^*p}(Q/Q_s(x))$
is exactly symmetric under the exchange $Q\leftrightarrow Q_s(x)$.
These two features have been observed in the HERA data 
\cite{Stasto:2000er}.

We would now like to refine the model by including
multiple gluon exchanges, which as we will argue, are
important when $Q\sim Q_s(x)$. Throughout we will preserve
the symmetry just pointed out.

\subsection{Accounting for multiple interactions}

\label{subsec:s2}

Usually multiple gluon exchanges
are disfavoured by the smallness of the coupling constant
and/or by powers of $Q_s^2(x)/Q^2$.
We can see this by computing the probability that
a dipole going through the proton undergoes an 
interaction.
If the dipoles are uniformly spread over the surface of the
proton, this probability is
\begin{equation}
p=\frac{1}{N}\sigma^{\gamma^*p}(x,Q^2)\cdot\frac{\Lambda^2}{\pi}
\simeq 2\pi\as^2\frac{Q_s^2}{\pi Q^2}
\end{equation}
as long as it is small enough compared to 1.
Note that $p$ coincides with the packing factor $\kappa$
usually defined in the context of saturation
(see for example \cite{Levin:2001eq}).

We first observe that parametrically $2\pi\as^2\!\sim\! 1$ 
for $\as$
evaluated at the scale $Q\!\sim\! Q_s\!\sim\! 1\ \mbox{GeV}$. 
Then when $Q_s^2(x)\!\ll\! Q^2$, $p$ is much less than 1:
multiple gluon exchanges are disfavoured by a power of this small factor.
When $Q^2\!\sim\! Q_s^2(x)$ however, $p$ is of order $1$ and
it is not justified to neglect multiple gluon exchanges.

In a classical framework, these multiple interactions could
be accounted for by modifying the interaction probability
in the following way.
We note that
$e^{-p}\equiv S^2$ is the probability of no interaction,
where $S$ is the $S$-matrix element for the dipole-proton
interaction for a given impact parameter (it is constant
over the area of the proton within our approximations).
The dipole-proton total cross section is then
$2\pi/\Lambda^2\cdot(1\!-\!S)$.
Multiplying it by the rate $N$ 
of splittings $\gamma^*\rightarrow q\bar q$,
we obtain for the total photon-proton cross section
\begin{equation}
\sigma^{\gamma^*p}(x,Q^2)=
N\cdot\frac{2\pi}{\Lambda^2}(1-S)
=\frac{2\pi N}{\Lambda^2}\left(1-e^{-p/2}\right)\ .
\end{equation}
The exponential is reminiscent of a Glauber resummation
which would be well justified 
for example in the case of a nuclear target. 
But we must keep in mind that the resummation is 
theoretically not
under control 
for a proton target
and there is up to now no systematic way to
do it in QCD.

We choose to keep the exponential form but we add a free
parameter $\nu_>$ in its argument, so that
we recover Eq.(\ref{eq:sansresom}) in the two-gluon exchange limit
and the strength of the 4-gluon interaction be fixed
by the data. We will check phenomenologically
that the result is not very
sensitive to $\nu_>$.
Absorbing all dimensionless normalizations
in the parameters $N$ and
$\nu_>$, we obtain the following formula: 
\begin{equation}
\sigma^{\gamma^*p}(x,Q^2)=N\frac{1}{\Lambda^2}\frac{1}{\nu_>}
\left\{1-\exp\left(-\nu_>\frac{Q_s^2(x)}{Q^2}
\left(1+\log\frac{Q}{Q_s(x)}\right)\right)\right\}\ .
\label{eq:finalehighq2}
\end{equation}
Note that this formula is very close to the original GBW model
(see Eq.(\ref{eq:gbwmod})). The main difference is the $\log$ present
in the exponential, which will play an important role
for the description of the data, as we
shall see in the following section.

Let us now consider the case $Q\!<\!Q_s(x)$.
We wish to preserve the symmetry of 
$Q/Q_s(x)\sigma^{\gamma^*p}(Q/Q_s(x))$
for $Q\leftrightarrow Q_s(x)$ at large $|\log(Q/Q_s(x))|$.
Hence we take the following ansatz for the 
cross section in this regime:
\begin{equation}
\sigma^{\gamma^*p}(x,Q^2)=N\frac{1}{\Lambda^2}
\frac{Q_s^2(x)}{Q^2}\frac{1}{\nu_<}
\left\{1-\exp\left(-\nu_<\frac{Q^2}{Q_s^2(x)}
\left(1+\log\frac{Q_s(x)}{Q}\right)\right)\right\}\ .
\label{eq:finalelowq2}
\end{equation}
We have kept the same parameters except $\nu_>$ which is
replaced by $\nu_<$. Indeed, there is no reason why the
resummation should be the same for $Q>Q_s(x)$ and for
$Q<Q_s(x)$.

Let us now try to interpret formula~(\ref{eq:finalelowq2}).
The probability that a specific dipole in the proton has
one interaction is 
$p=(\sigma^{\gamma^*p}(x,Q^2)/n)\cdot Q^2/(\pi N)$.
By interpreting the proton as a single effective dipole, with
flux $Q_s^2(x)/\Lambda^2$ and the photon as a set
of dipoles with density $N$, we are led to Eq.(\ref{eq:finalelowq2})
by the same argument of a Glauber resummation of multiple
gluon pair exchanges.
Of course, this is only an {\it a posteriori}
interpretation which has to be taken with great care.
Indeed, it is usually argued that for $Q<Q_s(x)$,
the dipoles in the proton are strongly correlated.

\section{A comparison to the data}
\label{sec:data}

In this section,
we compare Eq.(\ref{eq:finalehighq2}) (for $Q>Q_s(x)$) and
Eq.(\ref{eq:finalelowq2}) (for $Q<Q_s(x)$) to the 
recent ZEUS data for the $F_2$ structure function
both in the high \cite{Chekanov:2001qu} 
and low \cite{Breitweg:2000yn} 
$Q^2$ regime, and for $x\leq 10^{-2}$.
We assume the parametrization~(\ref{eq:qs}) for the
saturation scale $Q_s(x)$.
We end up with 5 free parameters to be determined
by the fit: $x_0$ and $\lambda$ parametrize $Q_s(x)$,
$N$ is the global normalization, and $\nu_>$ and
$\nu_<$ is the strength of the 4 gluon coupling
with respect to the 2 gluon coupling, for $Q>Q_s(x)$ and
$Q<Q_s(x)$ respectively.

The fit is performed with the 177 experimental points
which remain after the selection $x\leq 10^{-2}$
and $Q^2\leq 150\ \mbox{GeV}^2$.
The $\chi^2$ is 1.09 per degree of freedom.
It rises to 1.15 when we take all available $Q^2$
(185 data points).
The parameters for $Q_s(x)$ 
are $\lambda=0.35$ and $x_0=1.88\cdot 10^{-3}$.
The exponent for the rise of the parton densities $\lambda$ is
consistent with the one found within other approaches.
The saturation scale is 1~GeV already for relatively
high $x$ ($\sim 2\cdot 10^{-3}$), but we cannot make a point of
this fact
since many neglected effects could shift
the saturation scale by a constant factor (for instance
the full photon wave function would have such an effect).
The exponents $\nu_>$ and $\nu_<$ are very loosely determined
by the data: we find $\nu_>=0.8$ and $\nu_<=0.6$,
with an error of about $50\%$ and $100\%$ respectively.
If we fix them, to 1 for instance, the quality of the fit
does not drop much ($\chi^2=1.4/\mbox{d.o.f.}$).

The result of the fit is presented on Figs.~2 and~3.
We have added some of the available data points
which were not included in the fit and which
sit at large $Q^2$ and large $x$.
We see a good agreement over the whole $Q^2$, $x$
range. The $x$-slope seems to be reproduced fairly
well even at large $Q^2$. However, we slightly
overshoot the
normalization in this region.

We have also shown, for comparison, the cross section
without multiple scatterings, given by Eqs.(\ref{eq:sansresompert})
and~(\ref{eq:sansresom}). We observe that
the data is well reproduced for $Q\!\ll\! Q_s(x)$ and for
$Q\!\gg\! Q_s(x)$, but the resummations
are needed around the transition region 
$Q\!\sim\!Q_s(x)$.
We note that without the resummation 
of multiple gluon exchanges through the exponentiation,
we would obtain a very poor global fit. However, we have
a very good agreement with the data
when we fit separately formulae~(\ref{eq:sansresompert})
and~(\ref{eq:sansresom})
to the high and low $Q^2$
data, but with very different parameters $x_0$ and $\lambda$
in the two regions.

We plot the quantity ${Q}/{Q_s(x)}\sigma^{\gamma^*p}(x,Q^2)$
in Fig.~4. The symmetry under the exchange of $Q$ and $Q_s(x)$ is
apparent. Our parametrization reproduces it exactly,
by construction.
We see that the Glauber resummations
help to describe the early turnover
seen near $Q^2/Q_s^2(x)\!\sim\!3$.

A final remark concerns the comparison with the Golec-Biernat
and W\"usthoff model.
We have already noted that in the  high-$Q^2$
domain, the main difference between our
formula~(\ref{eq:finalehighq2}) and the GBW model is the $\log$ in
the exponent. We have checked that this term is crucial to
fit the data. In the GBW model of
Ref.~\cite{Golec-Biernat:1998js}, it is absent from the exponent,
but there is an overall $\log$ factor coming from the photon
wave function.

\section{Summary and discussion}

We can summarize our model as follows.
We have assumed that in the dipole frame the energy evolution of the
proton leads to the 
multiplication of the partons and consequently to the
appearance of a transverse momentum scale $Q_s(x)$.
When $Q\!>\!Q_s(x)$, we have the usual DIS picture, where the
photon probes a set of independent partons.
The rate of the growth of the parton densities has been parametrized
by $Q_s^2(x)/\Lambda^2$.
This procedure led us to formula~(\ref{eq:sansresompert}).
Then, we have taken seriously its extrapolation
to small values of $Q^2$, and we have observed that the symmetry
of the dipole-dipole cross section by exchange of the dipole
sizes is exactly seen in the data for $\gamma^*p$ total
cross section.
When $Q\!\sim\!Q_s(x)$, we have argued that the probability of
multiple interactions between the photon and the proton
becomes sizeable. A Glauber-like resummation was introduced
for $Q\!>\!Q_s(x)$ and extrapolated by symmetry for $Q\!<\!Q_s(x)$.
Although very simple and based on very
qualitative ideas, the model agrees quantitatively with
the recent high precision ZEUS data.

This model provides an intuitive picture of geometric scaling
both in the large $Q$ (hard Pomeron, perturbative)
and small $Q^2$ (soft Pomeron, non-perturbative
but weakly coupled) regimes based on
a symmetry
between low $Q^2$ and high $Q^2$ recently 
observed in the photon-proton cross section.

On the theoretical side,
we may observe that our model could be 
related
to unitary models based on Regge theory (see Ref.\cite{Capella:2000hq}).
In these models, a separation scale of 1~$\mbox{GeV}^{-1}$
between small and large distance is assumed. This could correspond
to our ``saturation'' radius $1/Q_s$ which is of the same order. 
In our approach, the fan diagrams involving triple Pomeron
vertices are hidden in the parametrization of the proton evolution
(they are explicitly present in Kovchegov's approach for instance).
The formula (\ref{eq:finalehighq2}) may be seen as
an eikonalization of fan diagrams.
On the other hand,
in a recent paper \cite{Mueller:2002zm}
the dipole-proton scattering amplitude
was derived in the context of linear BFKL supplemented by 
absorptive boundary conditions 
(see also \cite{Iancu:2002tr}).
The final formula looks very much like Eq.~(\ref{eq:sansresompert}).
It differs only by an anomalous dimension which we donot
take into account in our qualitative approach. 
The $\log$ has also presumably a different origin. 
However, this gives us confidence that the formula
we propose is anyway well rooted in QCD in the restricted
perturbative domain $Q\!>\!Q_s(x)$.

\section*{Acknowledgments}

I would like to thank E.~Iancu for his comments, L.~McLerran
for his interest,
K.~Golec-Biernat 
and R.~Peschanski
for their encouragements.


\clearpage

\noindent
{\large\bf Figure captions}\\

\vskip 1cm
\noindent
{\bf Figure 1.} {\em A picture of the photon-proton interaction in
the dipole frame}\\

In this frame, the momenta of the photon and of the proton
are collinear.
The photon
splits in a $q\bar q$ pair of size $1/Q$.
The QCD evolution
leads to a set of dipoles of sizes $1/Q_s(x)$ at the proton level
at the time of interaction. These two systems scatter
through the exchange of colour singlet pairs of gluons.
The cross section is depicted for
{\bf a.}~the exchange of two gluons,
{\bf b.}~the exchange of four gluons.

\vskip 0.8cm
\noindent
{\bf Figure 2.} {\em The $F_2$ structure function for medium $Q^2$}\\

Our parametrization of $F_2$ is shown (solid line)
together with the latest published ZEUS data
\cite{Chekanov:2001qu}.
The fit was done only for $x\!\leq\!10^{-2}$.
The dashed line is the simple dipole cross section
without taking account of many gluon exchanges (see formulae~
(\ref{eq:sansresompert}) and~(\ref{eq:sansresom})).
The errors shown for the data are the quadratic sum of the
statistic and systematic errors.\\

\vskip 0.8cm
\noindent
{\bf Figure 3.} {\em The $F_2$ structure function for low $Q^2$}\\

The same in the low $Q^2$ region.
The data points are from the ZEUS collaboration \cite{Breitweg:2000yn}.
We only show the $Q^2$-bins for which a significant
number of data points are available.

\vskip 0.8cm
\noindent
{\bf Figure 4.} {\em Geometric scaling}\\

The scaled photon-proton total cross-section
${Q}/{Q_s(x)}\sigma^{\gamma^*p}(x,Q^2)$
is plotted against the combined
variable $Q^2/Q_s^2(x)$. The data are from the ZEUS
collaboration
\cite{Chekanov:2001qu,Breitweg:2000yn}, 
and are selected according to $x\!\leq\!10^{-2}$.

\clearpage
\begin{center}
\epsfig{file=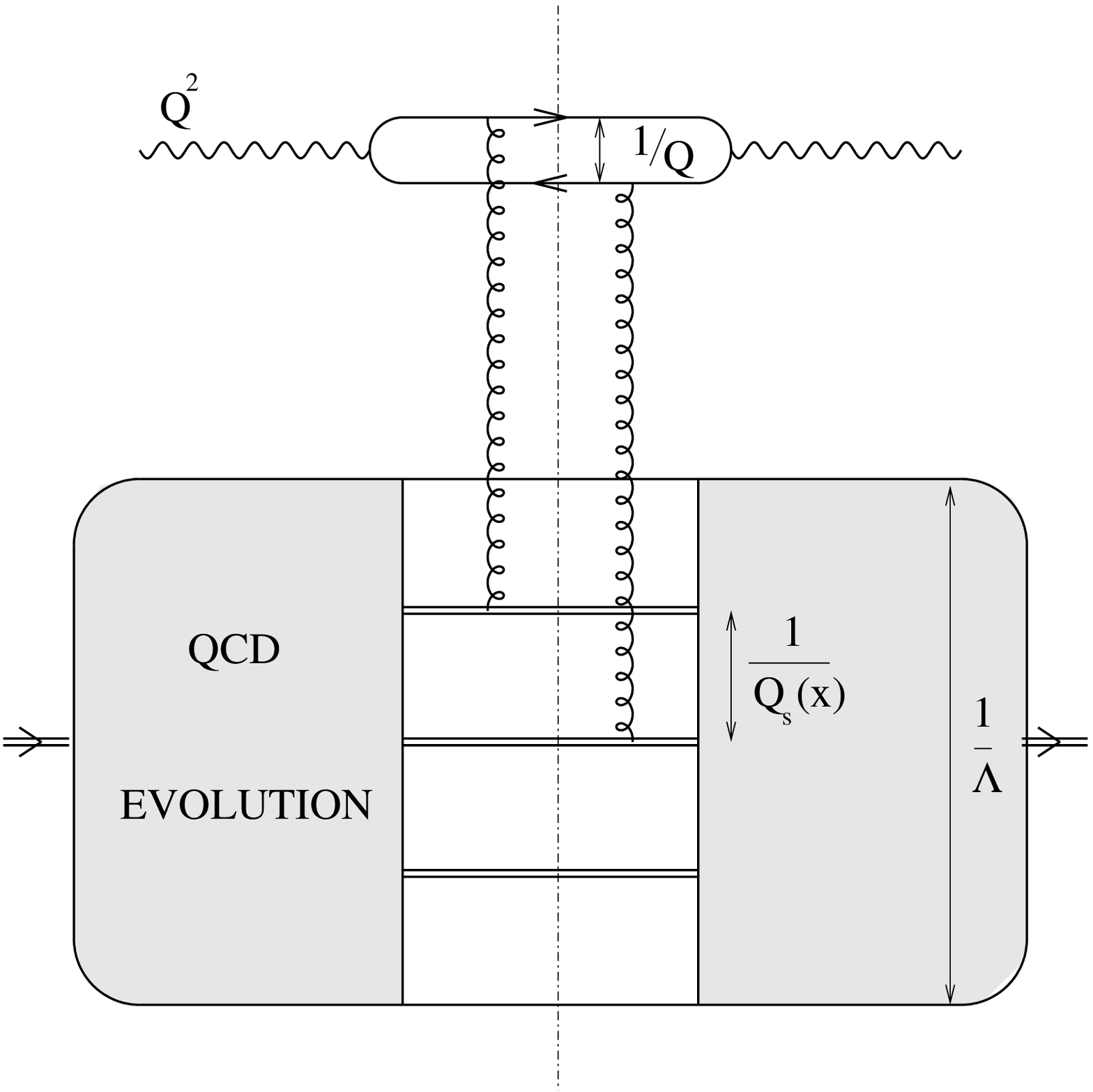,width=8cm}\\
\vskip 0.8cm
{\bf a.}\\
\vskip 2.cm
\epsfig{file=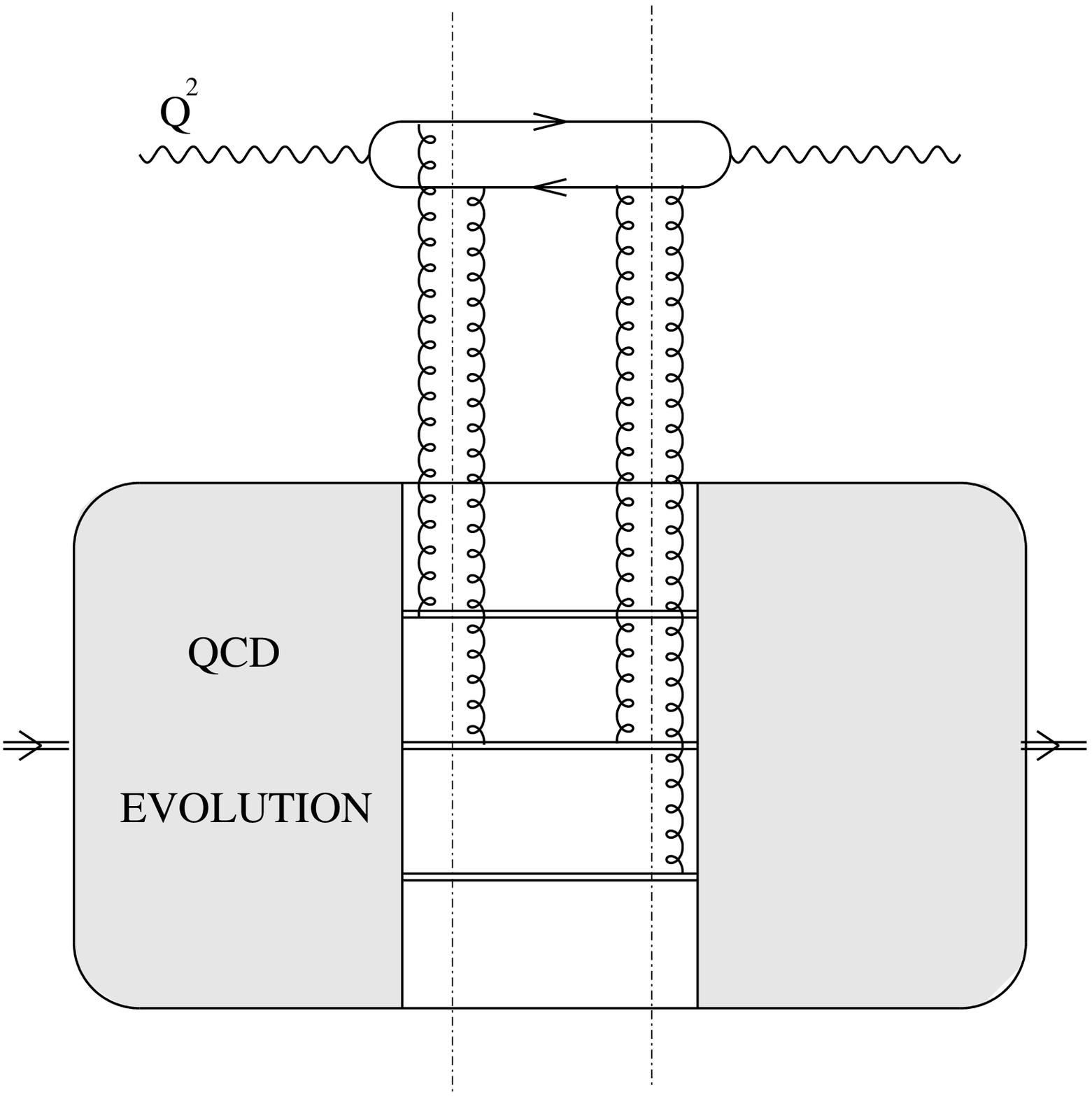,width=8cm}\\
\vskip 0.8cm
{\bf b.}\\
\end{center}
\vskip 0.5cm
\begin{center}
{\bf Figure 1}
\end{center}

\clearpage

\setlength{\oddsidemargin}{0.75cm}
\setlength{\textwidth}{19.5cm}

\begin{center}
\epsfig{file=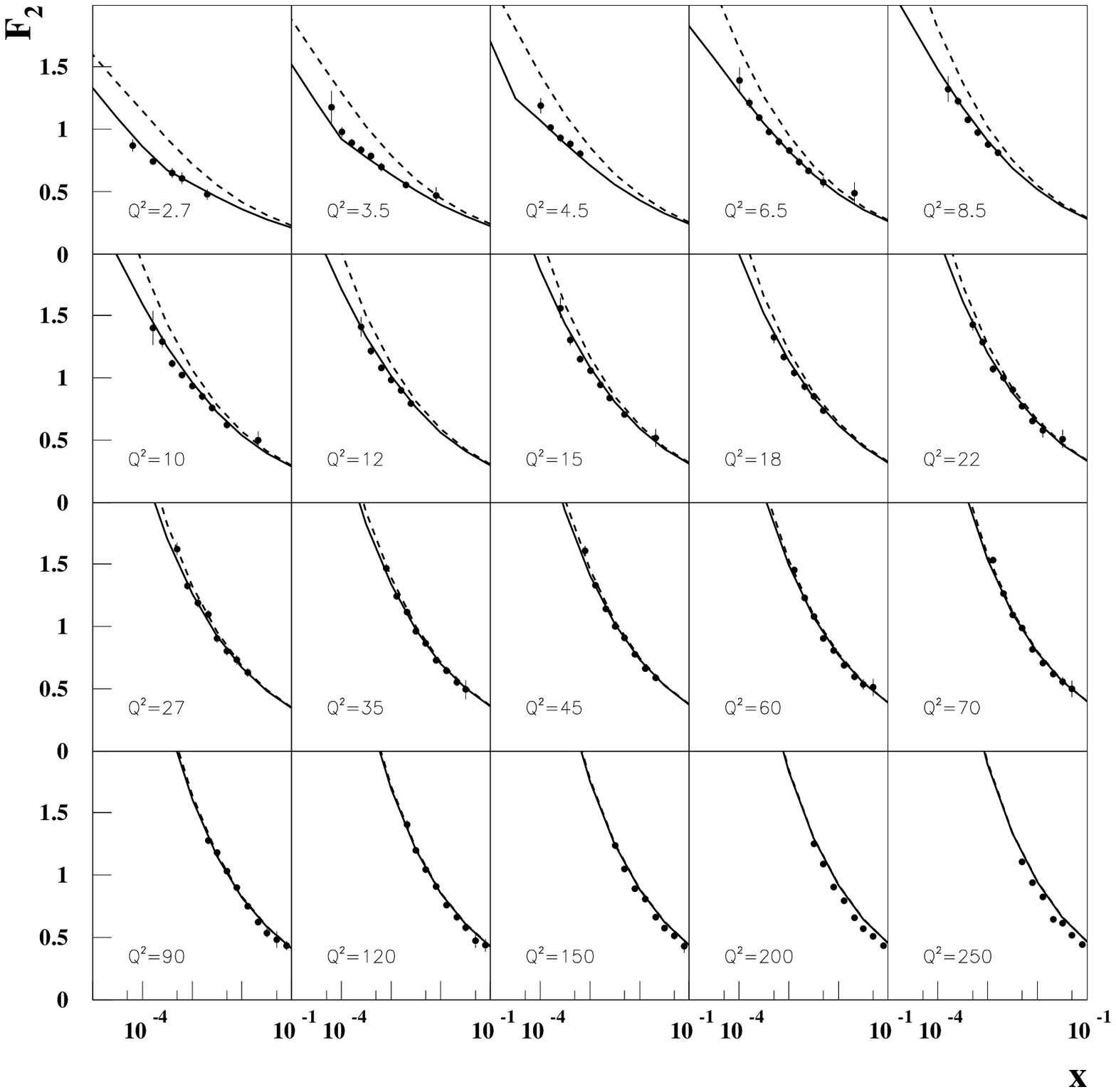,width=18cm}
\end{center}
\vskip 0.5cm
\begin{center}
{\bf Figure 2}
\end{center}

\clearpage
\begin{center}
\epsfig{file=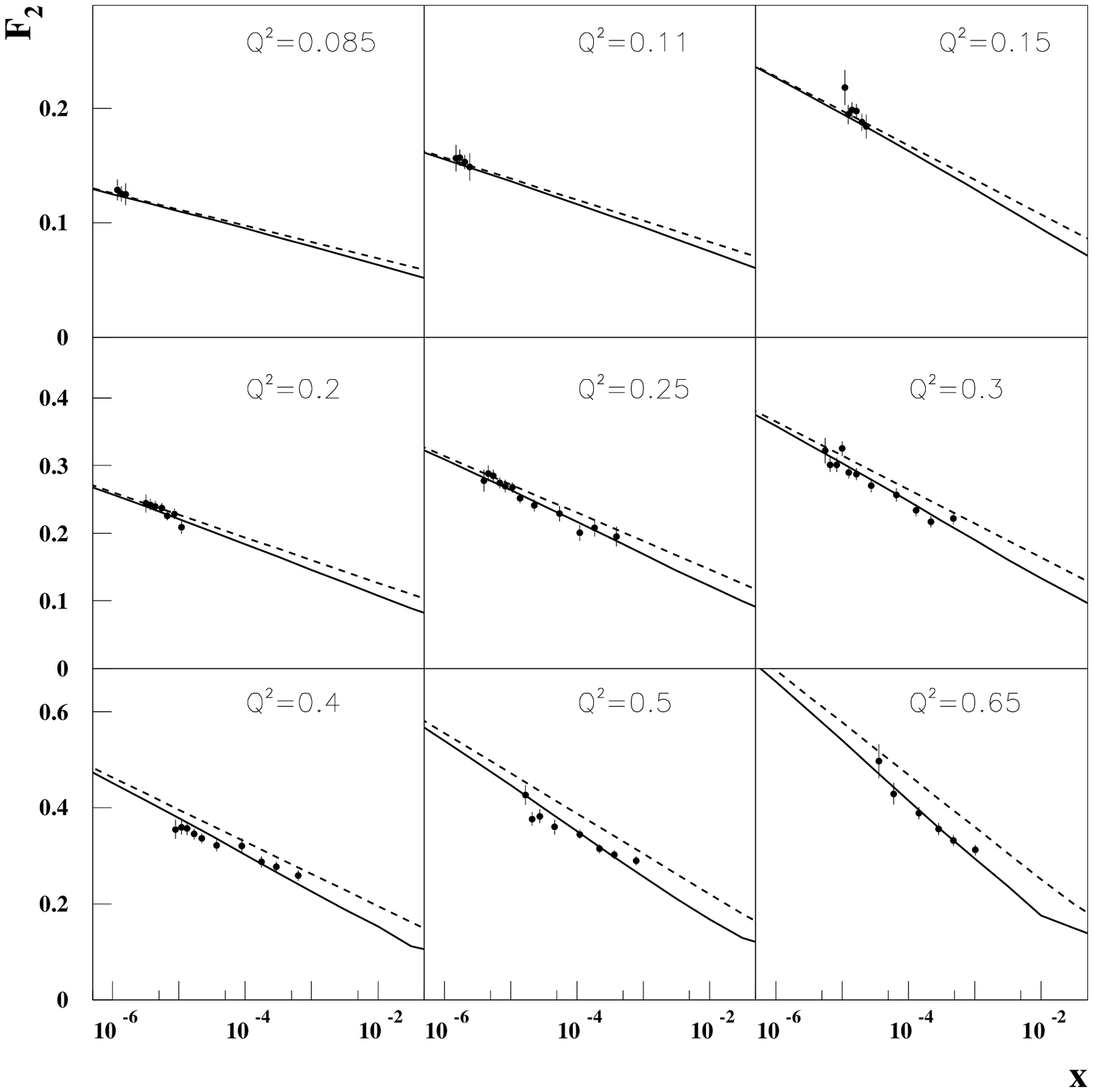,width=18cm}
\vskip 0.5cm
{\bf Figure 3}
\end{center}

\clearpage
\begin{center}
\epsfig{file=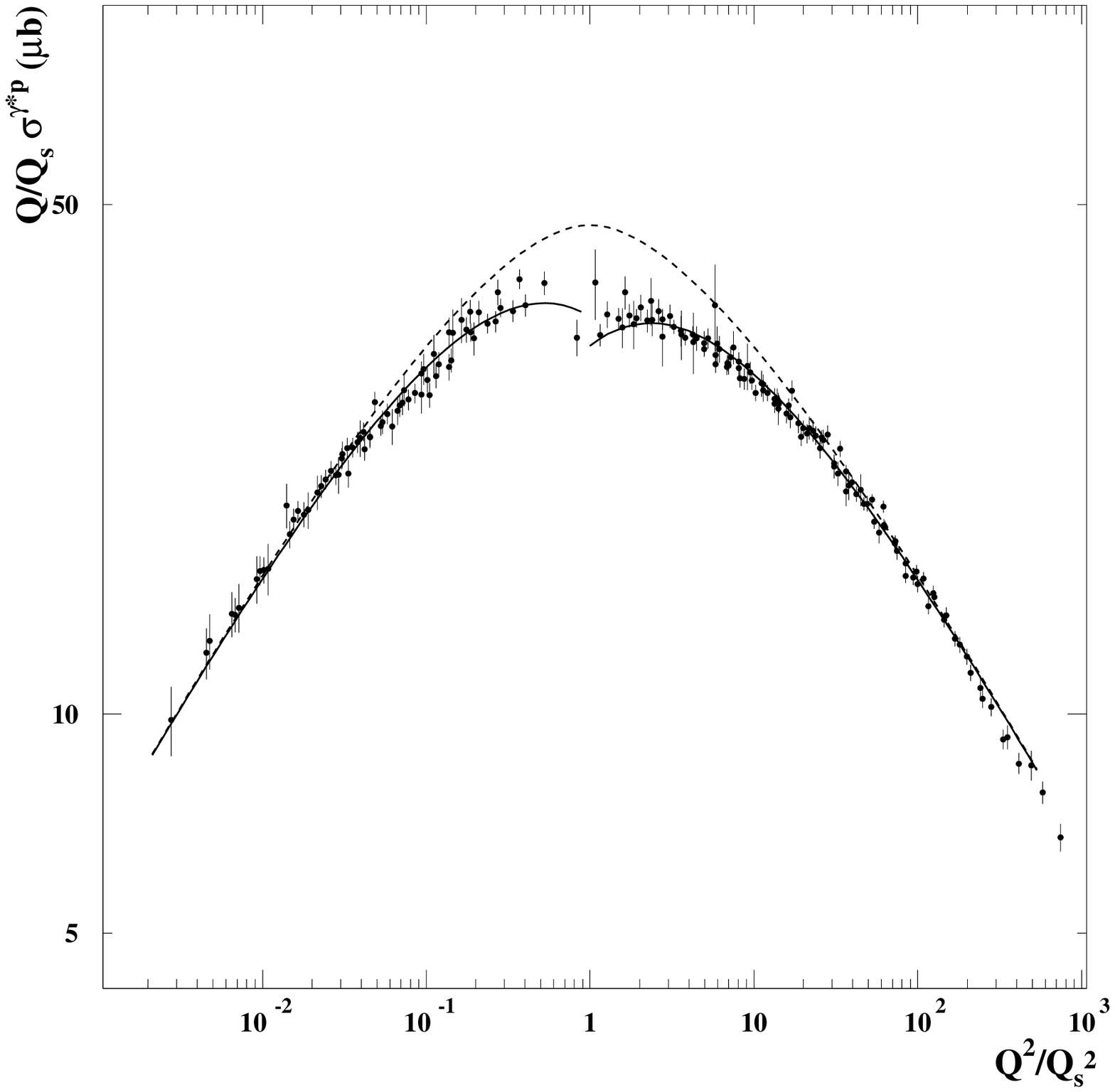,width=18cm}
\vskip 0.5cm
{\bf Figure 4}
\end{center}

\end{document}